\documentclass[JCAP,twocolumn,nofootinbib,superscriptaddress]{revtex4-1}
\input epsf
\usepackage{graphics}
\usepackage{amsmath,amssymb} 
\usepackage{color} 
\usepackage{dcolumn}
\usepackage{hyphenat}
\usepackage{multirow}

\usepackage{graphicx}

\def\be{\begin{equation}}
\def\ee{\end{equation}}
\def\ba{\begin{eqnarray}}
\def\ea{\end{eqnarray}}

\newcommand{\beqa}{\begin{eqnarray}}
\newcommand{\eeqa}{\end{eqnarray}}

\newcommand{\beq}{\begin{equation}}
\newcommand{\eeq}{\end{equation}}

\newlength{\tskip}
\newlength{\colwidth}

\begin{document}

\title{Primordial Black Hole Dark Matter and the LIGO/Virgo observations}

\author{Karsten Jedamzik} 
\affiliation{Laboratoire de Univers et Particules de Montpellier, UMR5299-CNRS, Universite de Montpellier, 34095 Montpellier, France}

\begin{abstract}
The LIGO/Virgo collaboration have by now detected the 
mergers of ten black hole binaries via the emission 
of gravitational radiation. The hypothesis
that these black holes have formed during the cosmic QCD epoch and make up
all of the cosmic dark matter, has been rejected by many authors reasoning
that, among other constraints, primordial black hole (PBH) dark matter would lead to
orders of magnitude larger merger rates than observed. We revisit the
calculation of the present PBH merger rate. Solar mass PBHs form clusters at fairly high redshifts, which evaporate at lower redshifts. 
We consider
in detail the evolution of binary properties in such clusters
due to three-body interactions between the two PBH binary members and a third by-passing PBH, for the first time, by full numerical integration. 
A Monte-Carlo analysis shows that formerly predicted merger rates are
reduced by orders of magnitude due to such interactions. The natural prediction
of PBH dark matter formed during the QCD epoch yields a pronounced peak
around $1M_{\odot}$ with a small mass fraction of PBHs on a shoulder around 
$30M_{\odot}$, dictated by the well-determined equation of state during the
QCD epoch. We employ this fact to make a tentative prediction of the
merger rate of $\sim 30M_{\odot}$ PBH binaries, and find it very close
to that determined by LIGO/Virgo. Furthermore we show that current LIGO/Virgo
limits on the existence of $\sim M_{\odot}$ binaries do not exclude QCD PBHs
to make up all of the cosmic dark matter. Neither do constraints on QCD PBHs
from the stochastic gravitational background, pre-recombination
accretion, or dwarf galaxies pose a problem. Microlensing constraints on
QCD PBHs should be re-investigated.
We caution, however, in
this numerically challenging problem some possibly 
relevant effects could not be treated.   
\end{abstract}

\maketitle
Five years ago the decades-long efforts of 
the LIGO collaboration have taken fruit and led to
the first unambiguous detection~\cite{TheLIGOScientific:2016qqj} of the gravitational wave signal of an inspiraling massive black hole binary. Since this pivotal moment a larger
number of further binary coalescence have been observed partially
in collaboration with Virgo~\cite{LIGOScientific:2018mvr}. 
Currently there are 66 detections of which
11 have been analysed. It came somewhat as a surprise that most of these
binaries were made of fairly massive $\sim 30M_{\odot}$ black holes
(see however ~\cite{Belczynski:2009xy,Belczynski:2010tb}).
As star formation scenarios do not necessarily predict this, the
theoretical community pondered the question if these black holes could be
primordial and constitute the dark matter.   

It has been known for long that mildly non-linear, horizon size overdensities could collapse and form an event horizon, i.e. a primordial black hole
(PBH, hereafter)~\cite{Zeldovich:1967,Hawking:1971ei,Carr:1975qj}
(for a review cf. to \cite{Khlopov:2008qy}).
When such collapse occurs during radiation domination in the early
Universe, the dynamics is characterized by a competition between self-gravity
and pressure forces~\cite{Jedamzik:1996mr,Jedamzik:1998hc}, and observes the physics of critical phenomena
~\cite{Niemeyer:1997mt,Niemeyer:1999ak,Musco:2004ak}.
When pre-existing energy density perturbations, such as believed to emerge
from inflationary scenarios, are feature-less and almost scale-invariant, as observed in CMBR satellite missions, the equation of state
during the PBH formation epoch plays a crucial role. It has been 
argued~\cite{Jedamzik:1996mr,Jedamzik:1998hc}
that PBH formation during the QCD epoch would be particularly
efficient due to a softening of the equation of state. At the time of this
realization, the QCD phase transition was believed to be of first order.
Fully general relativistic numerical simulations of PBH formation 
confirmed that PBHs form more easily during the QCD 
epoch~\cite{Jedamzik:1999am}, leading to
a pronounced peak of PBHs on the $\sim 1M_{\odot}$ scale.  
Though the simulations were performed under the assumption of a first order transition, it was argued in ~\cite{Jedamzik:1996mr},
that any softening of the equation state,
as for example during higher order transitions, or annihilation phases, such as the $e^+e^-$ annihilation, would lead to a preferred scale in the PBH mass
function. In the case of the latter this mass scale
is around $\sim 10^5-10^6M_{\odot}$. With the advancements of lattice gauge simulations
it was possible to derive the zero chemical potential QCD equation of state
with high precision~\cite{Borsanyi:2016ksw,Bhattacharya:2014ara}. 
This equation of state, was recently used in 
approximate analytic calculations to derive the 
putative PBH 
mass function~\cite{Byrnes:2018clq,Carr:2019kxo,Sobrinho:2020cco}.
This mass function indeed has a very well developed peak at $M\approx 1M_{\odot}$ and broader shoulder around 
$M\sim 30M_{\odot}$. It is tantalizing that this second scale
is close to the inferred BH binary masses by the Ligo/Virgo collaboration. 

There is a whole suite of observational constraints on the abundances of 
PBHs in the solar mass range, including the MACHO/Eros Milky Way microlensing constraints~\cite{Alcock:2000ph,Tisserand:2006zx},
constraints from the
CMBR due to PBH accretion~\cite{Carr:1981,Ricotti:2007au,Ali-Haimoud:2016mbv,
Serpico:2020ehh,Hutsi:2019hlw}, dynamical constraints,
such as constraints from the abundances of wide stellar binaries in the 
galactic halo~\cite{Yoo:2003fr,Quinn:2009zg,Monroy-Rodriguez:2014ula}, 
constraints from dwarf galaxies~\cite{Brandt:2016aco,Koushiappas:2017chw} (cf. to ~\cite{Carr:2020gox} for a review). These constraints have been recently amended by limits on the PBH mass
fraction contributing to dark matter, due to the observed BH binary coalescence
rates form LIGO/Virgo data and from the non-observation of
a gravitational stochastic wave background due to PBH 
mergers~\cite{Bird:2016dcv,Mandic:2016lcn,Clesse:2016ajp,
Wang:2016ana,TheLIGOScientific:2016dpb,
Raidal:2017mfl,Abbott:2017xzg,Bringmann:2018mxj,
Bartolo:2019zvb,Wang:2019kzb}. 
The verdict of the community~\cite{Sasaki:2016jop,
Ali-Haimoud:2017rtz,Ballesteros:2018swv,Bringmann:2018mxj,Raidal:2018bbj,
Vaskonen:2019jpv,DeLuca:2020qqa} 
is that solar mass PBHs can not constitute the bulk of the dark matter, mostly because they would greatly surpass the observed LIGO/Virgo rates.

In this paper, we show that this conclusion is with high
likelihood incorrect, we believe that
QCD formed PBHs may very well constitute all of the dark matter.
The findings of this paper, should stimulate intense research on
a numerically difficult problem. Limits imposed from the coalescing rate
rely on the correctness of the PBH binary semi-major and eccentricity
distribution. The initial distribution has been first calculated in 
~\cite{Nakamura:1997sm,Ioka:1998nz} subsequently re-calculated 
by several authors~\cite{Sasaki:2016jop,Ballesteros:2018swv,Raidal:2018bbj}, 
and its subsequent
evolution has been deemed to be too unimportant~\cite{Ali-Haimoud:2017rtz,Ballesteros:2018swv,Bringmann:2018mxj,Raidal:2018bbj,Vaskonen:2019jpv,DeLuca:2020qqa} to circumvent LIGO/Virgo
constraints.
We will show that this distribution, in fact, dramatically 
evolves between the first formation
of binaries and the present day, allowing for consistency with
current data.

PBH dark matter has the characteristics of perfect cold dark matter on large scales, however, there are two important differences when compared to particle
cold dark matter. First, PBH dark matter is formed infinitely cold as any
peculiar motions of density fluctuations should have inflated away
during inflation. When binaries then form from such initial conditions,
the distribution function $P(a,e)$ of binary semi-major axis $a$ and 
eccentricity $e$ is highly peaked at $e\approx 1$, 
reflecting very radial orbits. The binary coalescence time, due to the
emission of gravitational waves is given by
\begin{equation}
t_{gw} = \frac{3}{170}\frac{1}{(GM_{bh})^{3}}a^4\bigl(1-e^2\bigr)^{7/2}
\end{equation}
and becomes very short when $e$ approaches unity. Any perturbation of the
binary would be expected to typically lower $e$, to get to a more equilibrated,
unpeaked steady state in $P(a,e)$. This in turn
drastically increases $t_{gw}$~\cite{Sigurdsson:1994ju,Fregeau:2004if,Raidal:2018bbj,Vaskonen:2019jpv}. We will show that simple three-body
interactions between the two binary members and a third PBH indeed have
this effect. 

The second difference to particle dark matter pertains to the granularity
of PBH dark matter. As individual dark matter particles are fairly
massive, PBH dark matter has an additional isocurvature density 
fluctuation component on small scales simply due to small-number statistics
and the associated Poisson noise, i.e the fluctuations of the number
of PBHs in a given volume containing $N$ PBHs on average, is
$\delta (N) = \Delta N/N = 1/\sqrt{N}$. Those isocurvature fluctuations
lead to the early formation of PBH clusters at redshifts as high
as $\sim 1000$. These 
clusters have been observed in numerical simulations, with the smallest
clusters observed in ~\cite{Raidal:2018bbj} and a full range of clusters
observed in ~\cite{Inman:2019wvr}. Contrary to prior 
claims~\cite{Chisholm:2005vm,Chisholm:2011kn} the PBH isocurvature
fluctuations have been confirmed to be 
Poissonian~\cite{Ali-Haimoud:2018dau,Desjacques:2018wuu}.

\begin{figure}[htbp]
\centering
\includegraphics[width=0.48\textwidth]{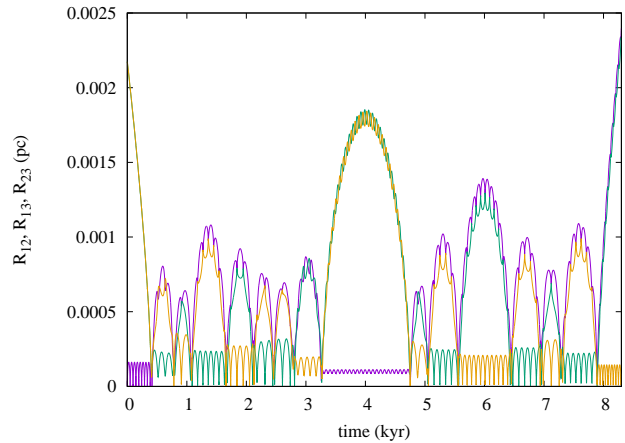}
\caption{\label{fig:fig3} An example of a scattering event.
Shown are the distances between the three PBHs
participating in the scattering event, $R_{12}$ (pink), the initial binary,
$R_{13}$ (green), distance between one member of the binary and the perturber,
and $R_{23}$ (yellow), distance between the other member and the perturber.
Initial binary properties: $a = 8.1\times 10^{-5}$pc,
$e = 0.999969$, $t_{gw} = 4.48\times 10^{-2}$Gyr. 
Final binary properties: 
$a = 7.59\times 10^{-5}$pc,
$e = 0.90$, $t_{gw} = 5.0\times 10^{10}$Gyr.
The impact parameter is $b=7.1\times 10^{-5}$pc.}
\end{figure}

\begin{figure}[htbp]
\centering
\includegraphics[width=0.48\textwidth]{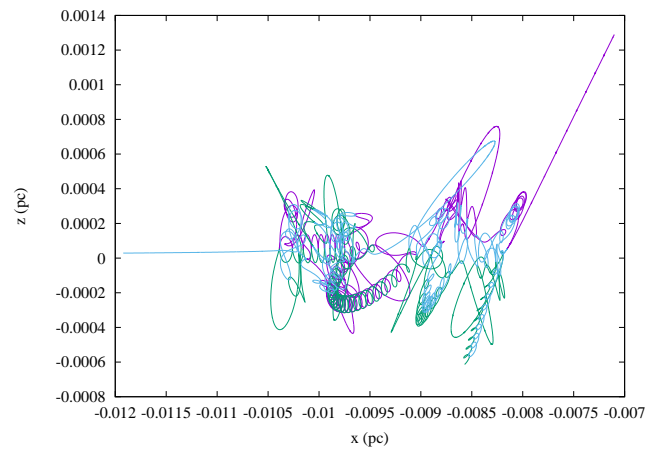}
\caption{\label{fig:fig4} The orbits of the three PBHs in the scattering event shown in Fig.1, in a particular plane in the initial binary CM frame. 
It is seen that the interaction is very complex.}
\end{figure}

Following ~\cite{Inman:2019wvr} the growth factor of isocurvature perturbations is given by
\begin{equation}
D(a) \approx \biggl(1+\frac{3}{2}\frac{a}{a_{eq}}\biggr)\, ,
\end{equation}
where $a$ is scale factor and $eq$ denotes matter-radiation equality.
Assuming a spherical top hat collapse model, we may require
$D(a)\delta (N) \approx 1.68$ to determine the collapse redshift
$a_{coll}$. In this model the final cluster density is given by
$\rho_{cl} \approx 178\langle \rho_{DM}\rangle (a_{coll})$ where
$\langle \rho_{DM}\rangle$ is the average dark matter density.
These relations allow us to determine the PBH number density
in clusters
\begin{equation}
n_{cl} \approx 1.6\times 10^{5}\frac{1}{\rm pc^3}
\biggl(\frac{M_{pbh}}{M_{\odot}}\biggr)^{-1}N_{cl}^{-3/2}
\end{equation}
and their virial velocity
\begin{equation}
v_{cl}\approx 0.60 \frac{\rm km}{\rm s} 
\biggl(\frac{M_{pbh}}{M_{\odot}}\biggr)^{1/3}N_{cl}^{1/12}
\end{equation}
as a function of the number of PBHs in the cluster $N_{cl}$ 
and PBH mass $M_{pbh}$. Their properties are shown in Table 1.

It is well known~\cite{BT08,Afshordi:2003zb} that dense clusters with a moderate number
of cluster members $N_{cl}$ are unstable towards complete evaporation
over the life-time of the Universe. Small clusters are therefore only of
temporary existence between the initial collapse redshift $z_{coll}$ and the evaporation redshift $z_{evap}$
The approximate time scale for complete evaporation (up to a last
remaining binary) of PBH clusters may be estimated~\cite{BT08} by 
\begin{equation}
t_{evap}\approx 140t_{relax}\approx 
14\frac{N_{cl}}{{\rm ln}N_{cl}}t_{cross}
\end{equation} 
where $t_{relax}$ is the cluster relaxation time and $t_{cross}\approx R_{cl}/v_{cl}$ is the cluster crossing time.
Here $R_{cl}$ can be determined by the relation 
$(4\pi/3)n_{cl}R_{cl}^3 = N_{cl}$.
Both the formation- and evaporation- redshifts,
$z_{form}$ and $z_{evap}$, are also shown in Table 1.

We have considered three-body interactions in these cluster 
environments. Since $v_{th}^{cl}\ll v_{b}^{int}$, where
$v_{th}^{cl}$ is the virial cluster velocity and
$v_{b}^{int}$ is the internal binary velocity, we may not use
the impact approximation to derive analytic results, but rather
have to resort to full numerical computations. 
A three-body scattering
event is characterized by a variety of parameters, 
binary member masses $M_{b1}$, $M_{b2}$ perturber mass $M_p$,
semi-major axis $a$, eccentricity $e$,
two inclination angles of the binary plane $\theta$ and $\phi$ with respect
to the perturber velocity direction, the
position of the binary members on the orbit given by a parameter $\psi$, impact parameter at infinity $b$, as well as the perturber velocity $V_p$ in 
the binary CM frame.  The accuracy of individual scattering
calculations is confirmed by surveilling energy and angular momentum
conservation.

\begin{figure}[htbp]
\centering
\includegraphics[width=0.48\textwidth]{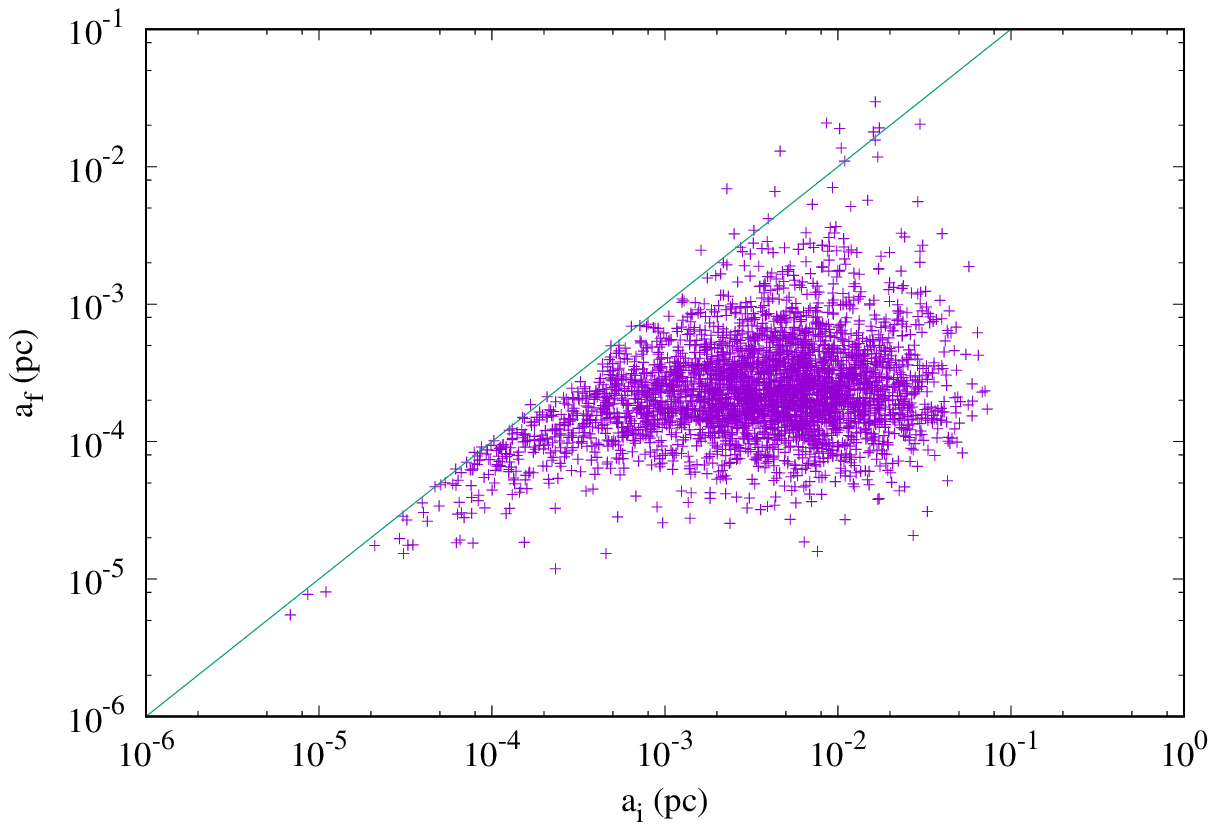}
\includegraphics[width=0.48\textwidth]{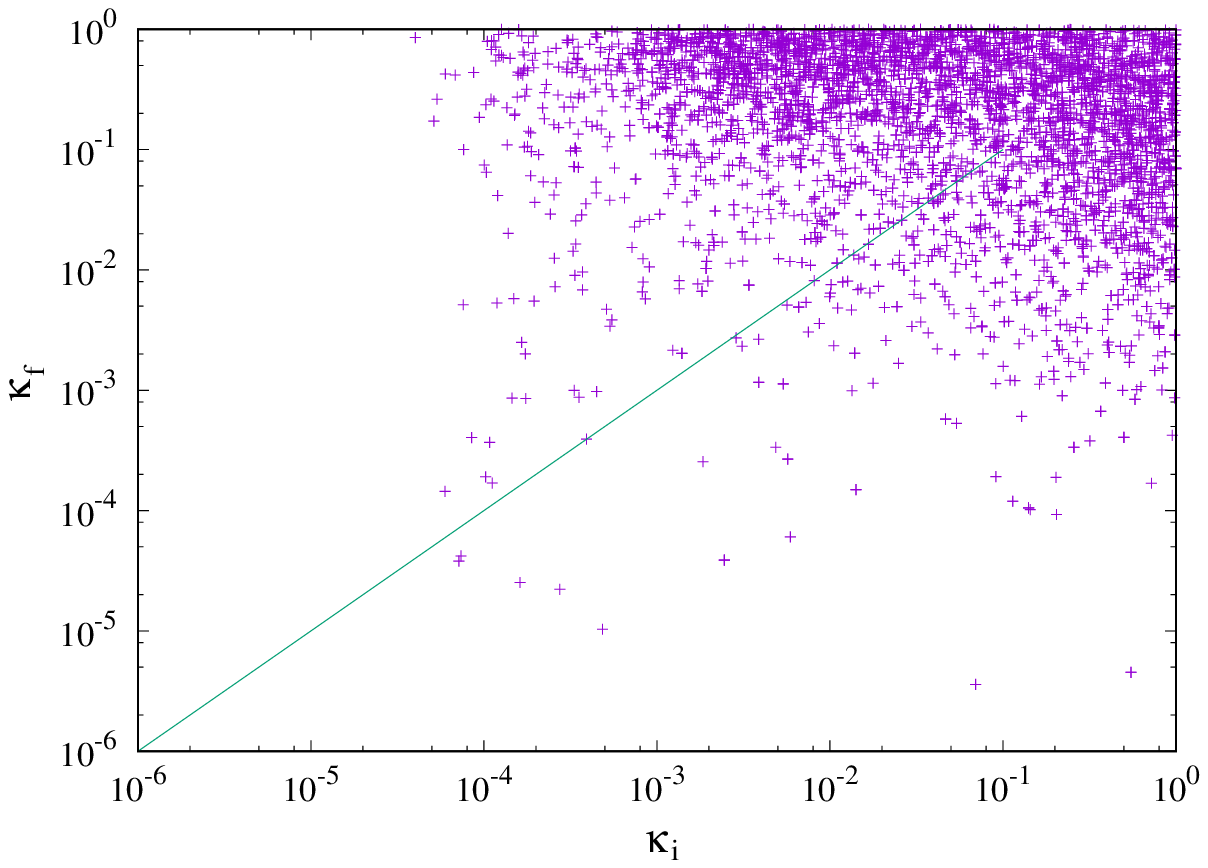}
\includegraphics[width=0.48\textwidth]{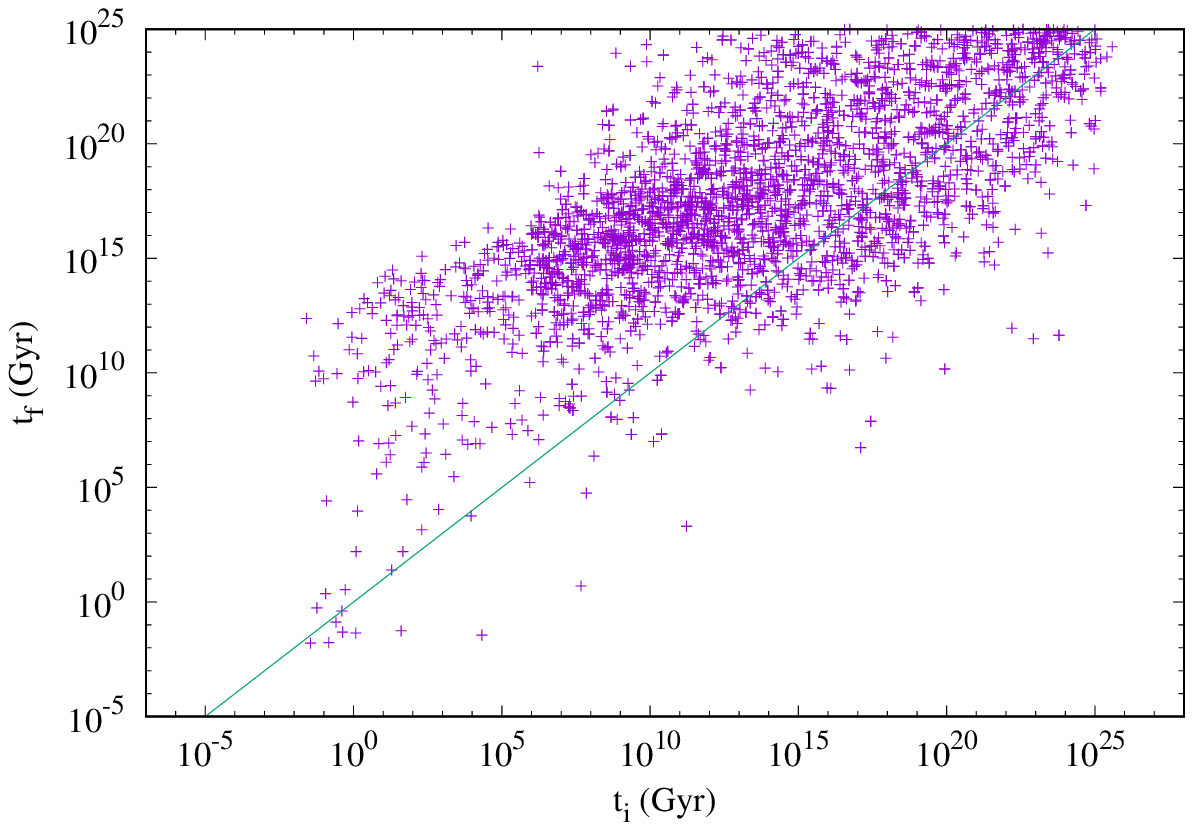}
\caption{\label{fig:cl} Scatter plots of a Monte Carlo simulation
of $2\times 10^4$ evolutions of binary properties due to three-body interactions
in a cluster with $N_{cl}=100$ and $M_{pbh} = M_{\odot}$. Each point shows
the final semi-major axis $a_f$ versus the initial $a_i$ (upper panel),
the final eccentricity  parameter $\kappa_f=(1-e_f^2)$, where $e$ is eccentricity, versus the initial $\kappa_i$ (middle panel), 
and the final merging time $t_{gw}^f$ versus the initial $t_{gw}^i$ (lower panel). The three lines $a_i=a_f$, $\kappa_i=\kappa_f$, and
$t_{gw}^i = t_{gw}^f$ are shown to guide the eye.}
\end{figure}

Fig.1 and Fig.2 show one example of such three-body scattering. 
This not atypical
scattering event on a hard and very eccentric binary illustrates the
complexity of such scatterings. Several times the binary changes partner,
binary $a$ and $e$ are constantly modified, until ultimately one PBH
escapes to infinity, leaving a binary which is even harder, in agreement
with the Heggie-Hills conjecture~\cite{Heggie:1975,Hills:1975}. However,
though the binary is harder, its eccentricity has changed
away from $e\approx 1$, implying that $t_{gw}$ has increased from
an initial $4.48\times 10^{-2}$Gyr to $t_{gw} = 5.0\times 10^{10}$Gyr.

We have performed Monte-Carlo (hereafter MC) simulations of the effects of three-body
scatterings on pre-existing binary properties due to their presence in
PBH clusters between redshifts $z_{form}$ and $z_{evap}$. 
The initial binary properties were taken from the distribution
calculated in ~\cite{Nakamura:1997sm,Sasaki:2016jop}.  The MC simulation considers the $10$ most
important scattering events, where importance is gauged by the smallness
of the impact parameter $b$. The simulation assumes $N_{cl}=100$, 
however, results for $N_{cl}=10$ and 
$N_{cl}=1300$ are very similar. For our simulations $M_{b1}=M_{b2}=M_p= M_{\odot}$ are assumed. Cluster frame binary and perturber velocities 
are given random directions and their magnitude is drawn from
a flat distribution ranging from $v_p=0$ to $v_p=v_{th}^{cl}$ for perturbers
and $v_b=0$ and $v_b=v_{th}^{cl}/\sqrt{2}$ for binaries. Here the factor of $\sqrt{2}$ takes into account that binaries are twice as heavy. 
The perturber velocity in the binary CM frame is subsequently obtained by a
Galilean transformation. The angles $\theta$, $\phi$, and $\psi$ are randomly
drawn from their appropriate distribution functions.
Results are shown in Figure 3.

\begin{table*}[!tbp]
\centering
\begin{tabular}{c|c|c|c|c|c|c|c|c|c|c|c}
 & $f_d$ & $f_{bcl}$ & $f_{bej}$ & $f_{dcl}$ & $z_{form}$ &  $z_{evap}$ & $n_{cl}$ ${\rm (1/pc^3)}$ & $v_{cl}$ $({\rm km/s})$ & $R_{cl}$ ${\rm (pc)}$ \\
\hline\hline
$N_{cl} = 10$ & 0.74 & 0.23 & 0.02 & 0.97  & 1200  & 165 & 9500 & 0.76  & 0.063  \\
$N_{cl} = 100$ & 0.78 &  0.21 & 0.005 & 0.97 & 320 & 15 & 190  & 0.93  & 0.50     \\
$N_{cl} = 1300$ & 0.81 & 0.18 & 0.003 & 0.98 & 85 & 0 & 3.6 & 1.1 & 4.4  \\
 \hline
\end{tabular}
\caption{\label{tab:fractions} Shown as a function of the
number $N_{cl}$ of PBHs in the cluster, assuming $M_{pbh}=1M_{\odot}$, are the fraction of PBH binaries which are destroyed by three-body interactions $f_d$, the fraction of binaries which stay binary
and stay in the cluster $f_{bcl}$,the fraction of PBH binaries which stay binary and are ejected
from the cluster $f_{bej}$, the fraction of PBHs which after binary
disruption stay in the cluster $f_{dcl}$, the formation redshift of the cluster, the
evaporation redshift of the cluster, the cluster density, the cluster 
virial velocity, and the cluster radius. }
\end{table*}

Figure 3 illustrates the following effects. All hard binaries,
i.e. binaries with $a {}^{<}_{\sim} 10^{-3}$pc are getting
harder due to three-body interactions, consistent with the Heggie-Hills conjecture~\cite{Heggie:1975,Hills:1975}. Eccentricities are
drastically lowered due to scatterings. For example, most of the
binaries with
$\kappa_i \sim 10^{-4}$, corresponding to eccentricities 
$e_i\sim 0.99995$ are scattered to values $\kappa_f {}^{>}_{\sim} 10^{-2}$,
corresponding to $e_f < 0.995$. Here $\kappa$ is defined as
$\kappa = (1-e^2)$. By this effect $t_{gw}$ increases 
by a factor $> 10^7$. It is seen from the lowest panel
in Fig.3 that most binaries with $t_{gw}{}^{<}_{\sim} 10^{13}$Gyr get scattered
to $t_{gw}{}^{>}_{\sim} 10^{13}$Gyr. 
This is likely due to an equilibration of binary properties due to
the complexity of multiple binary-perturber interactions as the one shown in the Fig. 1 and Fig. 2.
Fig. 3 illustrates that any conclusion of limits on PBH dark matter
relying on the distributions computed in ~\cite{Nakamura:1997sm,Sasaki:2016jop} is highly suspect.
Many authors~\cite{Clesse:2016ajp,Wang:2016ana,Raidal:2017mfl,
Ali-Haimoud:2017rtz,Byrnes:2018clq,Ballesteros:2018swv,Authors:2019qbw,
Vaskonen:2019jpv,DeLuca:2020qqa} have used this distribution to draw conclusions.

Due to PBH three-body interactions, i.e. close encounters of binaries
with single PBHs, many of the binaries do get destroyed. Table 1 lists the
fraction of binaries which get destroyed in PBH clusters. It also
lists the fraction of PBHs which stay in the cluster after destruction.
We observe in our MC simulations that only the wider binaries get destroyed, with the hard binaries staying intact. It is also observed that a large
fraction of the binaries stay in the cluster. We note here that our 
criteria for ejection from the cluster is that
the post-scattering velocity is larger than the escape velocity, assumed
to be $v_{esc}^{cl} = 2 v_{th}^{cl}$~\cite{BT08}. The internal energy of a binary is given by $E_b = -GM_{b1}M_{b2}/2a$.
As a binary gets destroyed, the cluster kinetic energies of the PBHs
are reduced by this amount. This reduction of total cluster kinetic
energy will lead to the shrinking of the cluster.
Hard binaries get harder during scatterings, i.e. $a$ reduces.
In this case, the kinetic energies of perturber and binary are substantially
larger after the three-body interaction leading to an expansion of the
cluster. It is likely that the first effect
dominates, due to the larger number of wide binaries. The probable
shrinking of clusters is not taken into account in our study.

\begin{figure}[htbp]
\centering
\includegraphics[width=0.48\textwidth]{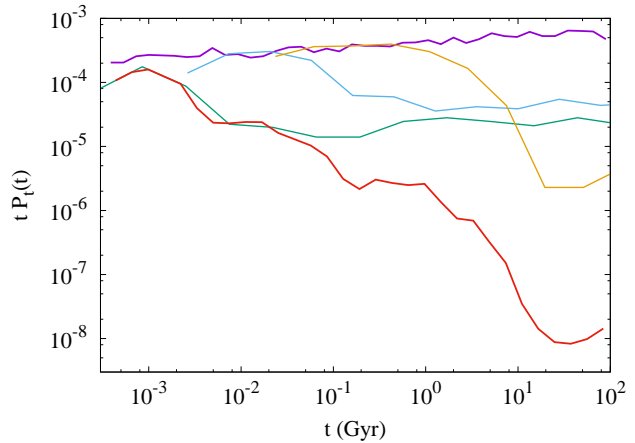}
\caption{\label{fig:destruction} The fraction of PBH binaries
$t\,{\rm d}P_t(t)/{\rm d}t$ which merge at time $t$, for all binaries
having $t_i< 100\,$Gyr. Initial distribution (pink)~\cite{Nakamura:1997sm,Sasaki:2016jop} and final
distributions after processing between $\tau_{form}$ and $\tau_{evap}$ 
in $N_{cl}=10$ (green), $N_{cl}=100$ (blue), 
and $N_{cl}=1300$ (orange) clusters are shown. The total
$t\,{\rm d}P_t(t)/{\rm d}t$ after successive membership in 
$N_{cl}=10$, $N_{cl}=100$, and $N_{cl}=1300$ clusters is 
shown by the red line.
It is seen that ${\rm d}P_t(t)/{\rm d}t$ at  $t = 14\,$Gyr
is reduced by a factor $\sim 3\times 10^{-5}$
with respect to its initial value. (cf. to the appendix concerning further
details on the production of this figure).
}
\end{figure} 

We have computed the evolution of the probability per unit time that a binary merges at time $t$, i.e. 
${\rm d}P_t(t)/{\rm d}t$ (cf. Eq.(13) of ~\cite{Sasaki:2016jop}).
The problem is numerically challenging
because one needs to obtain the statistics of tails, where already a
single scattering is difficult to calculate due to the mismatch of time
scales, $\tau_b\ll \tau_{cl}$, where $\tau_b$ is the binary orbital 
period and $\tau_{cl}$ is the typical PBH cluster crossing time. The tail
one is interested in are the very few evolutions
where a binary gets scattered into, or stays at, the small coalescence time
$t_0\approx 14\,$Gyr, the current cosmological time. 
We were not able to solve the complete problem, due to numerical restrictions,
but only computed the evolution of ${\rm d}P_t(t)/{\rm d}t$ 
for all PBH binaries
having $t_i < 100\,$Gyr. Results are shown in Fig. 4. 
Shown is the change of $t\,{\rm d}P_t(t)/{\rm d}t$ from its initial value due to
evolution in $N_{cl}=10,100$ and 1300 clusters, during the time interval
$\tau_{form}(N_{cl})$ and $\tau_{evap}(N_{cl})$. 

Fig. 4 illustrates that PBH clusters of widely differing $N_{cl}$ have
in magnitude similar effects on ${\rm d}P_t(t)/{\rm d}t$.
One may have expected less evolution in larger clusters as their densities
are low. However, this effect is almost canceled by the longer evaporation
time of large clusters. The number of collisons over a given time interval
$\Delta t$ is 
\begin{equation}
\Delta N_{sc} = \pi b^2 v_{cl} n_{cl} \Delta t
\end{equation}
where $b$ is impact parameter. Treating only the $N_{sc}$ most important 
events for each cluster, one may show with the help of the equations above
that
\begin{equation}
b_{typical}(N_{cl}) \propto \frac{{\rm ln^{1/2}} N_{cl}}{N_{cl}^{1/6}}
\label{eq:z}
\end{equation}
is almost independent of $N_{cl}$, with the rhs of Eq.~\ref{eq:z}
being $1.03$,$1.00$, and $0.81$ for $N_{cl}=10$, $N_{cl}=100$, and
$N_{cl}=1300$, respectively. In this way binaries receive major 
perturbations in three different environments, over strongly increasing
time intervals and in strongly decreasing densities.
We note here that for the initial conditions of the $N_{cl}=100$, and
$N_{cl}=1300$ simulations, strictly speaking we should not use the results
of ~\cite{Nakamura:1997sm,Sasaki:2016jop}, but rather the distribution
of $a$ and $e$ obtained from processing in the smaller cluster of the
previous generation. Our simulations are currently not sophisticated enough 
to do so.

It is important to realize
that in the spirit of hierarchical structure formation, $N_{cl}=10$ clusters
will at lower redshifts be incorporated in $N_{cl}=100$ clusters, which in 
turn, at even lower redshift will become part of a $N_{cl}=1300$ cluster. Binaries which had been ejected or evaporated from a smaller cluster will
quickly thermalize by three-body interactions in the larger cluster, and may
not escape the potential well of the larger cluster, unless they get ejected
again by another catastrophic collision. It is therefore that
the total reduction of ${\rm d}P_t(t)/{\rm d}t$ is multiplicative,
being the product of reduction rates in all three different size cluster
environments. The total $t\,{\rm d}P_t(t)/{\rm d}t$ is shown by the red line, illustrating a total
factor $\sim 3\times 10^{-5}$ reduction at $t_0 \approx 14\,{\rm Gyr}$. This 
reduction may be directly applied to the observed merger rates as those are
given by ${\cal M} = n_{pbh}^{avg}{\rm d}P_t(t_0)/{\rm d}t$, with 
$n_{pbh}^{avg}$ the average present day cosmic PBH number density.

When taking a scenario of PBHs formed during the QCD epoch in its most
natural form, i.e. without any pre-tuned peak of inflationary perturbations
on a particular scale, but only taking into account effects of the 
softening of the equation of state~\cite{Jedamzik:1996mr,Jedamzik:1998hc}, 
most PBHs form on the $M_{\odot}$
scale, whereas only a mass fraction of $\sim 10^{-2}$ form on the larger
$\sim 30M_{\odot}$ scale~\cite{Byrnes:2018clq,Carr:2019kxo,Sobrinho:2020cco}. This estimate assumes Gaussian statistics.
In this case the relevant limits on
PBH dark matter from LIGO/Virgo come from two LIGO/Virgo analysis,
limits on $M_{\odot}$ PBH dark matter by the non-observation of 
solar mass PBH binaries~\cite{Authors:2019qbw} and the estimated merger
rate of BH-BH binaries on larger mass 
scales~\cite{LIGOScientific:2018mvr,LIGOScientific:2018jsj}. 
Using the main result of this paper given in Fig. 4, we may compute the
expected merger rates of pre-existing $\sim M_{\odot}$ PBHs binaries
to be ${\cal M}_{M_{\odot}}\approx 40\,{\rm Gpc}^{-3}{\rm yr}^{-1}$, to
be compared to the LIGO/Virgo imposed upper limit of
${\cal M}_{M_{\odot}}^<\approx 5.2\times 10^3\,{\rm Gpc}^{-3}{\rm yr}^{-1}$
~\cite{Authors:2019qbw}. For detailed
merger rates the reader is referred to our
companion paper~\cite{Jedamzik:2020omx}.
In fact, pre-existing binary properties get so efficiently modified by
collisions in clusters, that the total merging rate is dominated by
merging of single PBHs colliding at very low impact parameter~\cite{Mouri:2002mc,Bird:2016dcv,Clesse:2016ajp} 
in present-day clusters. We derive a rate of 
${\cal M}_{M_{\odot}}^d \approx 640 \,{\rm Gpc}^{-3}{\rm yr}^{-1}$ for
$N_{cl}=1300$ clusters and ${\cal M}_{M_{\odot}}^d \approx 160 \,{\rm Gpc}^{-3}{\rm yr}^{-1}$ for $N_{cl}=3000$ clusters~\cite{Jedamzik:2020omx}, 
well below the 
LIGO/Virgo limit.
Estimating ${\cal M}_{30M_{\odot}}$ is a bit more
tricky, as not even the initial distribution is known yet, and the
evolution of ${\rm d}P_t(t)/{\rm d}t$ again is rendered numerically unfeasible due to the small number $3\times 10^{-4}$ of $\sim 30M_{\odot}$ per solar mass
PBH. Obviously the evolution of ${\rm d}P_t(t)/{\rm d}t$ of $\sim 30M_{\odot}$ PBHs has to take into account the evolution of many more solar mass black holes. It would be expected that scatterings of $30M_{\odot}$ PBHs on 
$M_{\odot}-M_{\odot}$ or $30M_{\odot}-M_{\odot}$ binaries may typically lead to the exchange of the lighter black hole by the heavier one. 
This may well be the most important contribution to the abundance of 
$30M_{\odot} - 30M_{\odot}$ binaries. In any case, for a tentative estimate,
we assume here that as
with $M_{\odot}\,$ PBHs the contribution of pre-existing binaries is 
negligible. When considering the direct merger rate in present day clusters,
one has to take into account that $\sim 30\,M_{\odot}$ sink to the
center of the cluster. We estimate a merger rate of 
${\cal M}_{30M_{\odot}}^d \approx 12-150 \,{\rm Gpc}^{-3}{\rm yr}^{-1}$~\cite{Jedamzik:2020omx}
to be compared to the by
LIGO/Virgo inferred rate of $53.2^{+58.5}_{-28.8}\,{\rm Gpc}^{-3}{\rm yr}^{-1}$~\cite{LIGOScientific:2018jsj} at $90\%$ confidence level.

These results are rather promising. However, we caution that we were
not able to include several possibly relevant effects into the calculations. A
question of utmost importance, is which fraction of PBHs never enter into a cluster. Leaning on results of CDM simulations~\cite{Stucker:2017nmi}, this fraction could be 
rather small, but this has to be firmly established. 
In addition PBHs start infinitely cold which further could reduce
this fraction compared to cold dark matter. Further important effects are, the neglected scattering
of $t_i\gg 10\,$Gyr binaries into the $t_f\approx 10\,$Gyr band,
which could dominate destruction, as there are many more (i.e. $\sim 100$)
$t_i\, {}^>_{\sim}\, 10\,$Gyr
binaries than $t_i \, {}^<_{\sim}\, 10\,$Gyr, the shrinking of 
clusters due to binary destruction, and the role of core 
collapse of clusters~\cite{Vaskonen:2019jpv}. Finally, the inclusion of heavy PBHs into the sea of light ones
could also be of some importance. A complete treatment of this problem is
currently not straightforward due to a hierarchy of time scales 
$\tau_b\ll\tau_{cl}\ll\tau_H$, all of which need to be resolved, 
where $\tau_H$ is the Hubble scale. This may
only be achieved, if at all, by elaborate adaptive mesh techniques.
 
It is somewhat unfortunate, that the stellar mass scale and that for 
QCD era produced PBH dark matter are essentially the same. This fact
leads one to easily dismiss a potential discovery of dark matter, by
attributing such observations to particularities of star formation. After all,
we know stars exist. On the other hand, it is fortunate since gravitational wave
detectors optimized to observe binaries on the stellar scale, become at the
same time exquisite detectors for PBH dark matter formed during the
QCD epoch. In case LIGO/Virgo (in combination with optical observations) detect only one binary including a 
sub-Chandrasekar mass black hole, $M_{BH} {}^<_{\sim} 1.4M_{\odot}$, or
in a weaker version, one binary including a BH of mass in the by star formation forbidden gap $2M_{\odot} {}^<_{\sim} M_{BH} {}^<_{\sim} 5M_{\odot}$, the existence of PBH dark matter may have been
established.

In summary, the hypothesis that PBHs in the solar mass range 
make up the entirety of the dark matter had been seemingly ruled
out by a number of observational constraints.
However, taking our study at face value, even given the inherent 
uncertainties, the following statements can be made
\begin{itemize}
\item Current observational constraints from the merger rates of $\sim M_{\odot}$ PBHs
are evaded due to PBH collisions in clusters studied here.
\item
Similarly, observational constraints from the stochastic gravitational wave background due to PBH mergers should be evaded by the effect studied here.
\item Observational constraints from the CMBR due to accretion or the
dynamics of dwarf galaxies are evaded
if only a small fraction of the dark matter is in $\sim 30M_{\odot}$ PBHs,
as naturally predicted by the QCD equation of state.
\item The tentative estimate of the merger rate of $\sim 30 M_{\odot}$
PBHs is close to that observed by LIGO/Virgo.
\item The mass scale of $\sim 30 M_{\odot}$ PBHs is predicted by the QCD equation of state.  
\end{itemize}
Observational constraints from micro-lensing could still pose
a problem, but they should be re-investigated in the context of PBH clusters. 
We believe that the possibility that LIGO/Virgo 
has detected the dark matter should be taken seriously, and demands
further scrutiny.
\vskip 0.15in

{\it Acknowledgments.} I acknowledge detailed comments by 
Yacine Ali-Hamoud and Krzysztof Belczynski after first submission of
this document.
This paper was produced during Covid19 lockdown,
allowing for un-interrupted work.
I acknowledge my wife Nanci for being patient with me.
\vskip 0.15in

\appendix
\section{The Monte-Carlo simulations leading to Fig.4}

For the production of Fig. 4 simulations with $N_{cl}=10, 100\,$ and 1300
have been performed. For each cluster size of the order of $\sim 5000$
binaries were evolved. Here the $N_{cl}=10\,$ and 100 simulations consider 
the $10$ most important scattering events, whereas the simulation of 
$N_{cl}=1300$ clusters, treat the $\langle N_{sc}\rangle =30$ 
most important events. In the latter we had to switch to 
a larger $\langle N_{sc}\rangle$
since otherwise spurious effects of Poisson statistics would arise. Given the
long lifetime of such clusters, and the fact that we use Poisson statistics
for the time when an event occurs, the probability to have not 
undergone any scattering before $t$ is given by
${\rm exp}(-\langle N_{sc}\rangle t/t_0)$ which can be considerable at 
$t\sim 3-5\,$Gyr for low $\langle N_{sc}\rangle$, compared to the tail we want to compute. Each scattering
event was limited to five minutes of CPU time, and abandoned if taking longer.
Thus around $15\%$ of the to-be-evolved binaries were abandoned. These cases,
typically very hard binaries,
would subsequently be re-simulated with $\langle N_{sc}=4 \rangle$ 
and fifteen minutes of
CPU time limit. Here the reduction in $\langle N_{sc} \rangle$ 
helps as it reduces $z$,
and since for small $a/z$ simulations become very difficult. After this
second round of integration $0.8\%,\, 0.4\%,$ and $0.05\%$ for 
$N_{cl}=10,\, 100,$ and $1300$ clusters, respectively, had to be abandoned.
These abandoned runs are included in the figure with their initial
values. All of those abandoned runs were very hard binaries 
$a\sim 10^{-5}$pc at eccentricities very close to unity, with $t_{gw}$
substantially smaller than the present cosmic time.


%

\end{document}